\documentclass[aps,preprint,prb,superscriptaddress,showpacs,preprintnumbers,amsmath,amssymb]{revtex4-1}

\usepackage{amsmath}
\usepackage{amsfonts}
\usepackage{amssymb}
\usepackage{amsthm}
\usepackage{mathtools}

\usepackage{epsf}
\usepackage{graphicx}
\usepackage{epstopdf}

\usepackage{color}

\begin{document}

\title{Magnetic-field effect in the heterostructure Ba$_{0.8}$Sr$_{0.2}$TiO$_3$/LaMnO$_3$}

\author{D.~P. Pavlov}
\affiliation{Zavoisky Physical-Technical Institute, FRC Kazan
Scientific Center of RAS, 420029 Kazan, Russia}


\author{N.~N. Garif'yanov}
\affiliation{Zavoisky Physical-Technical Institute, FRC Kazan
Scientific Center of RAS, 420029 Kazan, Russia}

\author{A.~V. Leontyev}
\affiliation{Zavoisky Physical-Technical Institute, FRC Kazan
Scientific Center of RAS, 420029 Kazan, Russia}

\author{T.~M. Salikhov}
\affiliation{Zavoisky Physical-Technical Institute, FRC Kazan
Scientific Center of RAS, 420029 Kazan, Russia}

\author{V.~M.~Mukhortov}
\affiliation{Southern Scientific Center of RAS, 344006
Rostov-on-Don, Russia}

\author{A.~M.~Balbashov}
\affiliation{Moscow Power Engineering Institute, 111250
Moscow, Russia}

\author{R.~F. Mamin}
\affiliation{Zavoisky Physical-Technical Institute, FRC Kazan
Scientific Center of RAS, 420029 Kazan, Russia}

\author{V.~V.~Kabanov}
\affiliation{Zavoisky Physical-Technical Institute, FRC Kazan
Scientific Center of RAS, 420029 Kazan, Russia}
\affiliation{Department for Complex Matter, Jozef Stefan Institute,
1000 Ljubljana, Slovenia}

\date{\today}

\begin{abstract}
We have studied transport and magnetotransport properties of the
heterostructure consisting of the ferroelectric
Ba$_{0.8}$Sr$_{0.2}$TiO$_3$ film deposited on the single crystalline
manganite LaMnO$_3$. Two-dimensional electron gas arising at the
interface transforms the interface region of the antiferromagnetic
manganite to the ferromagnetic state with the reoriented magnetic
moments. We obtained that the contribution of these ferromagnetic
regions to the resistance of the heterostructure appears to be
dependent on the external magnetic field applied perpendicularly to
the interface (and parallel to {\it c}-axis of the ferromagnetic state of LaMnO$_3$).
We believe that the decrease of the resistance of the studied heterostructure under the influence of the external magnetic field is caused by an appearance of the ferromagnetic order in the interface area. This leads to a relatively small resistance. In turn, the cycling of the applied magnetic field leads to the increase of the regions of the ferromagnetic order and, as a result, to decreasing scattering of the current carriers at the ferromagnetic disordering.

\end{abstract}




\maketitle


The last few decades the problem of an interface between two layers
with competing properties such as nonmagnetic oxides with
ferroelectric oxide, antiferromagnetic oxide with ferroelectric
oxide, attracts considerable attention (see, e.g, \cite{Bozovic}). A
high mobility two-dimensional electron gas (2DEG) was first observed
at the interface between LaAlO$_3$ and SrTiO$_3$ in 2004
\cite{Ohtomo}. Antagonistic electronic properties of contacting
materials leads to unusual properties of the interface including the
appearance of 2DEG gas or even superconductivity \cite{Th,Rey,Bri}
and considerable changes in a magnetic state within few monolayers
in the vicinity of the interface.
The 2DEG was also found at the interface between magnetically
ordered Mott insulators and, in particular, ferromagnetic
(RE)TiO$_3$ with RE=Gd \cite{Moetakef}, Sm \cite{Jackson} and La
\cite{Biscaras} with highest density of the current carriers.  Later
on, the 2DEG formation was found in Nd(Pr)AlO$_3$  \cite{Annadi} and
LaGdO$_3$/SrTiO$_3$ \cite{Perna} heterostructures.
The 2DEG states were also reported at interfaces
between other nonmagnetic insulators, e.g., KTaO$_3$/SrTiO$_3$
\cite{Kal} and CaZrO$_3$/SrTiO$_3$ \cite{Chen}.

It is commonly believed that the local polarity of (LaO)$^{+1}$ and
(AlO$_2$)$^{-1}$ layers play an important role in the formation of
the 2DEG in the LaAlO$_3$/SrTiO$_3$ heterostructure. Bearing that in mind, we have chosen
Ba$_{0.8}$Sr$_{0.2}$TiO$_3$/LaMnO$_3$ heterostructure for our study.
Ba$_{0.8}$Sr$_{0.2}$TiO$_3$ exhibits the ferroelectric polarization
caused by the shift of Ti$^{+4}$ ions from the center of the oxygen
octahedron and the LaMnO$_3$ manganite is an antiferromagnetic
insulator.

In our previous paper \cite{Pavlov} we have studied the temperature
dependence of the resistivity of the heterostructure with the
ferroelectric Ba$_{0.8}$Sr$_{0.2}$TiO$_3$ film on the surface of the
single crystalline LaMnO$_3$ with the axis of polarization of
ferroelectric perpendicular to the surface of the single crystalline
LaMnO$_3$. The electrical resistance decreases significantly with
the lowering of the temperature below 165 K indicating the metallic
behaviour. The metallic behaviour of the resistance was observed only in samples where polarization of ferroelectric film was perpendicular to the interface\cite{Pavlov}. This indicates that indeed the abrupt change of the perpendicular to the interface component of polarization causes the formation of the 2DEG at the interface. Therefore, we can expect that the appearance of the charge carriers at
the interface with LaMnO$_3$ gives rise to the local ferromagnetic
state \cite{Dagotto,Gennes,Zener}. Note, that the creation of the highly conducting and even superconducting layer on the interface between ferroelectric Ba$_{0.8}$Sr$_{0.2}$TiO$_3$ film and antiferromagnetic La$_2$CuO$_4$ recently has been demonstrated  
in Ref. \cite{Pavlov2}. 

It is well known that manganites undergo the transition to metallic state with doping. For example, when strontium (Sr) is substituted to LaMnO$_3$
instead of La, the samples demonstrate the metallic behavior at relatively low
temperatures with a strontium concentration more than 17\%.
Essentially, the metallic state is ferromagnetic. The ferromagnetism
is caused by an indirect ferromagnetic exchange through the current
carriers. Therefore, when free carriers appear at the interface
leading to a metallic state, we expect that this state will have
ferromagnetic properties. This occurs due to a
strong ferromagnetic interaction through free current carriers with 
high concentration at the interface layer. 
The ferromagnetic properties of this 2DEG should be sensitive to the applied magnetic field.

Thus, the goal of our work is to detect the magnetoactive properties
of the conductive states at the interface. 
Since the interface layer is very
thin, we expect that the measurements of the ferromagnetic
properties of the interface against the background LaMnO$_3$ bulk
substrate is a difficult task. Therefore, it is better to use
samples when the Ba$_{0.8}$Sr$_{0.2}$TiO$_3$/LaMnO$_3$
heterostructure is realized in the form of sequentially deposited
LaMnO$_3$ and Ba$_{0.8}$Sr$_{0.2}$TiO$_3$ films. This is a project
for further advance of our research. In the present studies, we used samples that we have, in which a Ba$_{0.8}$Sr$_{0.2}$TiO$_3$ film was
deposited on top of a LaMnO$_3$ single crystal. We study the effect
of a magnetic field on the temperature dependence of the resistance of these samples.
We are primarily interested in how the magnetic field manifests
itself at low temperatures, because the low temperature
properties of the heterostructure are most strongly modified by the existence of the 2DEG at the interface.

Usually researches in this field are performed using the deposition
process controlled with very high accuracy, i.e. at a level of few
percent of an atomic monolayer or better. These heterostructures
have atomically flat and clean surfaces and interfaces. In the
present work we have studied in details the magnetic field and temperature
dependence of the electrical resistance of the heterostructure
which contains the ferroelectric Ba$_{0.8}$Sr$_{0.2}$TiO$_3$ film
deposited on the ab surface of the single crystalline manganite LaMnO$_3$. In accordance with the AFM measurements the surface of the single crystalline LaMnO$_3$ has
roughness of the order of 2 nm on 200-300 nm distance before the Ba$_{0.8}$Sr$_{0.2}$TiO$_3$ film was deposited. Because we
observe the strong suppression of the resistance at low
temperatures in our heterostructure therefore, the requirement of the high-quality interface is not absolutely necessary.
We would like also to underline that using a
ferroelectric material as an upper layer of the heterostructure
opens the possibility to control the properties of heterostructures
by switching the polarization in the ferroelectric layer.


Single crystalline substrates of antiferromagnetic LaMnO$_3$ were
prepared by  floating-zone crystal growth method. The samples were
mechanically polished. Then they were placed into the sputtering
chamber where they were covered by 350 nm thick ferroelectrical
Ba$_{0.8}$Sr$_{0.2}$TiO$_3$ layer. Before the deposition process
of Ba$_{0.8}$Sr$_{0.2}$TiO$_3$ on the top of LaMnO$_3$ single
crystal, the LaMnO$_3$  surface was cleaned with an argon plasma etching.
The thin film of Ba$_{0.8}$Sr$_{0.2}$TiO$_3$ was deposited in oxygen
atmosphere at 650 C and partial pressure of oxygen of 0.7 Torr by rf sputtering of polycrystalline  target of corresponding composition\cite{Mukhortov1,Mukhortov2,Mukhortov3}. The X-ray diffraction studies of our heterostructure have demonstrated that the ferroelectric BSTO film is epitaxial with $c$-axis of spontaneous polarization directed perpendicular to the interface. Note that the critical temperature of the ferroelectric transition T$_C \approx$ 540 K \cite{Mukhortov1} which corresponds to the ferroelectric transition temperature for thick epitaxial BSTO films.

The electrical resistance measurements were performed using $dc$
current in a standard four-terminal configuration. In order to measure the current and the voltage, 50 micron golden leads were attached to the surface of the
sample using silver paint.
%
%
The temperature of the sample was controlled
with the Cu-CuFe thermocouple.


Fig.~1 shows the temperature dependence of the resistance of
the Ba$_{0.8}$Sr$_{0.2}$TiO$_3$/LaMnO$_3$ heterostructure. In contrast
to the temperature dependence of the resistance for the single
crystalline manganite where the semiconducting behavior of the resistance is observed,
the resistance of the heterostructure is decreasing with decreasing temperature demonstrating a metallic below 165 K.
\begin{figure}[h]
\includegraphics[width=0.9\columnwidth,angle=0]{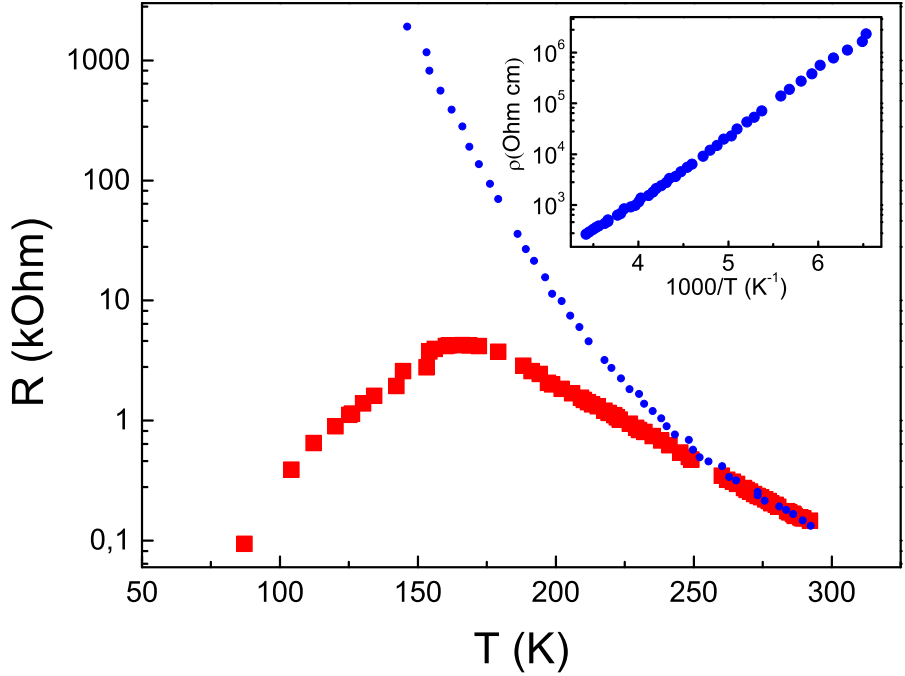}
\caption{The temperature dependence of resistance of
Ba$_{0.8}$Sr$_{0.2}$TiO$_3$/LaMnO$_3$ heterostructure
(shown in red squares). Blue circles in the figure
and in the inset show the temperature dependence of the resistance of
the single crystalline LaMnO$_3$.}
\end{figure}
Inset to Fig.~1 shows the result of the measurements of the
resistivity of the single crystalline LaMnO$_3$ substrate. The
measured temperature dependence of the resistivity demonstrates the
typical semiconducting behavior with the energy gap $\Delta_g \sim$0.3
eV which coincides well with the published data
\cite{Dagotto,Mukhin,Mukhin2}.

The most interesting feature of our results is the
accumulation effect of the magnetic field.
\begin{figure}[h]
\includegraphics[width=0.9\columnwidth,angle=0]{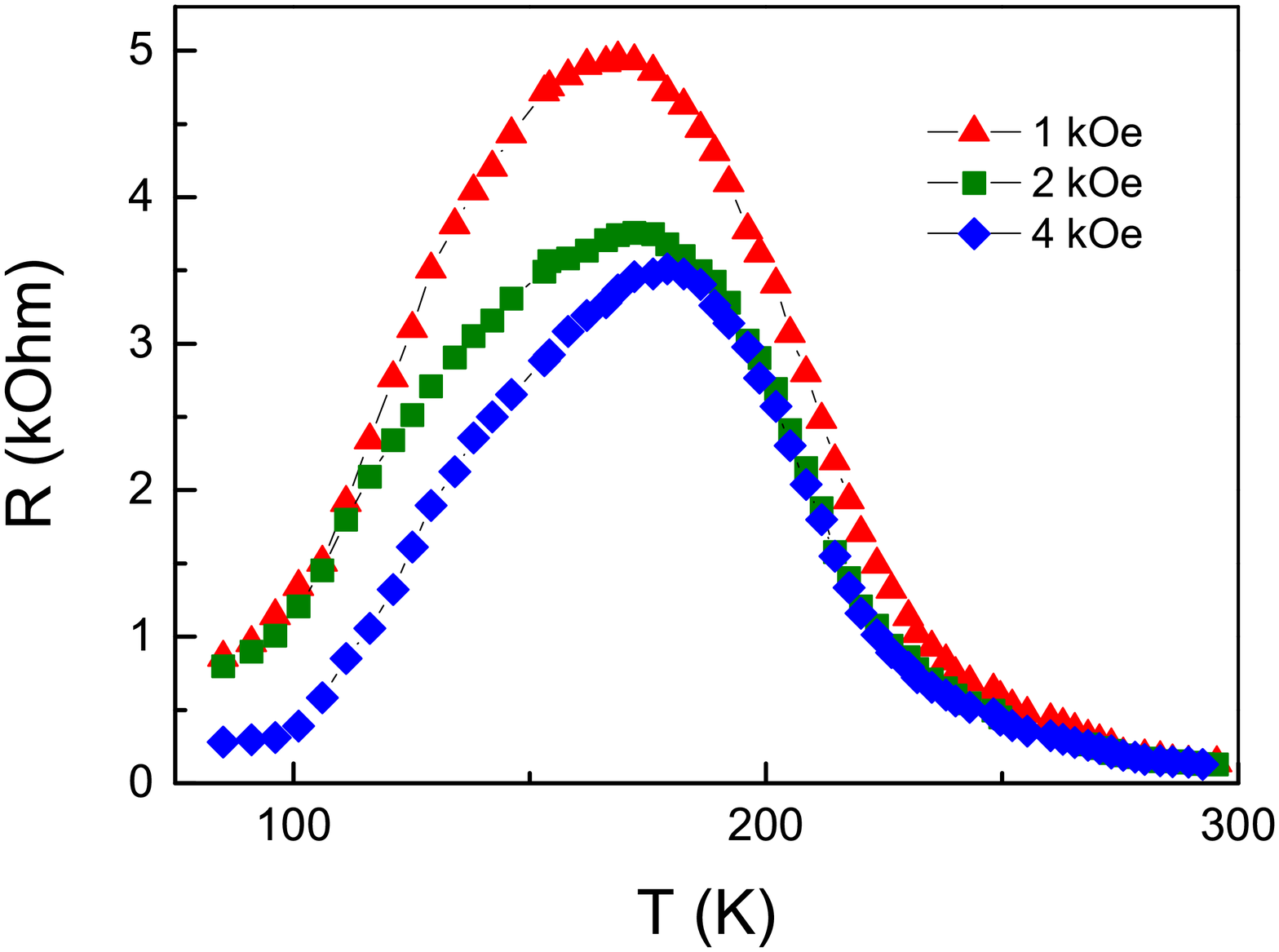}
\caption{The temperature dependence of resistance of the
Ba$_{0.8}$Sr$_{0.2}$TiO$_3$/LaMnO$_3$
heterostructure for magnetic field values shown in legend.}
\end{figure}
Fig.~2 demonstrates the first cycle of the measurements of the
temperature dependence of the resistance for different values of the
external magnetic field. As it is seen from Fig.2, the maximum in
the temperature dependence of the resistance decreases with increasing
of the magnetic field and shifts to the high-temperature region. The
next cycle starts from zero field with the value of the resistance
coinciding with the final value of the resistance in the previous
cycle (3 k$\Omega$). In the third cycle the maximum of the
resistance decreases to approximately 1 kOhm, and so on.
Finally, the accumulation effect of the magnetic field leads to the
magnitude of the maximum of the resistance of the order of 0.7
kOhm. The final result of the transformation of the temperature
dependence of the resistance under the influence of the external
magnetic field is
\begin{figure}[h]
\includegraphics[width=0.9\columnwidth,angle=0]{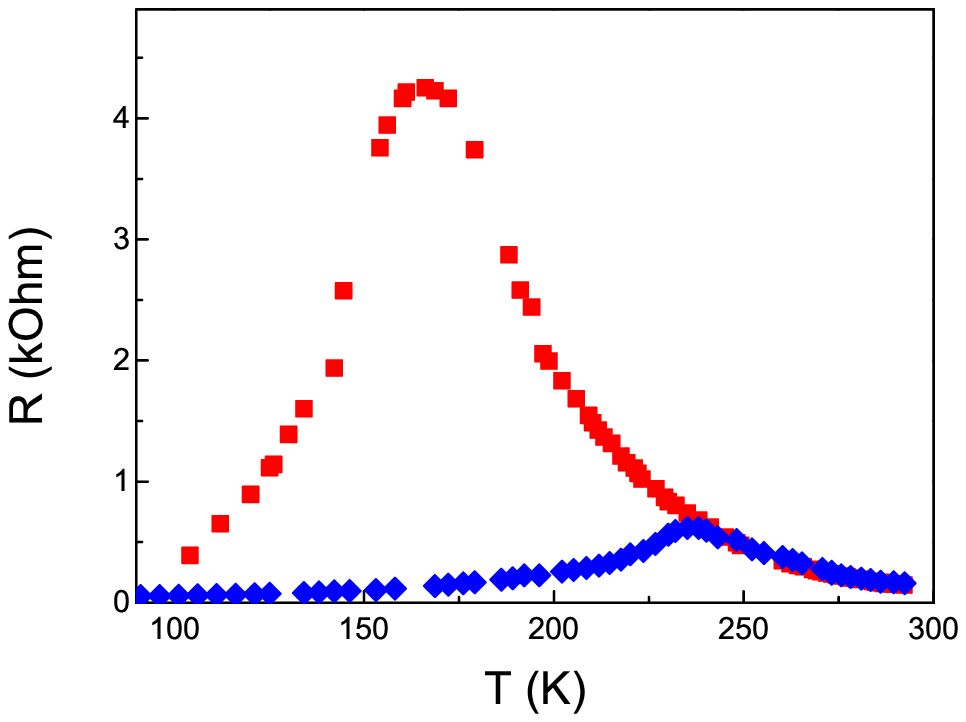}
\caption{The temperature dependence of resistance of the
Ba$_{0.8}$Sr$_{0.2}$TiO$_3$/LaMnO$_3$ heterostructure from Fig.~1
(solid squares). The final result of the transformation of the
temperature dependence of resistance of the studied heterostructure
(solid diamonds).}
\end{figure}
shown in Fig.~3. As it is seen from Fig.~3, the cycling of the
external magnetic field influences the resistance very strongly. The
magnitude of the maximum of the resistance decrease from 4.5
kOhm down to 0.7 kOhm. At the same time the position of
the maximum shifts from 170 to 230 K.

Further study of the magnetic field effect on R(T) curves is shown in Fig.~4.
\begin{figure}[h]
\includegraphics[width=0.9\columnwidth,angle=0]{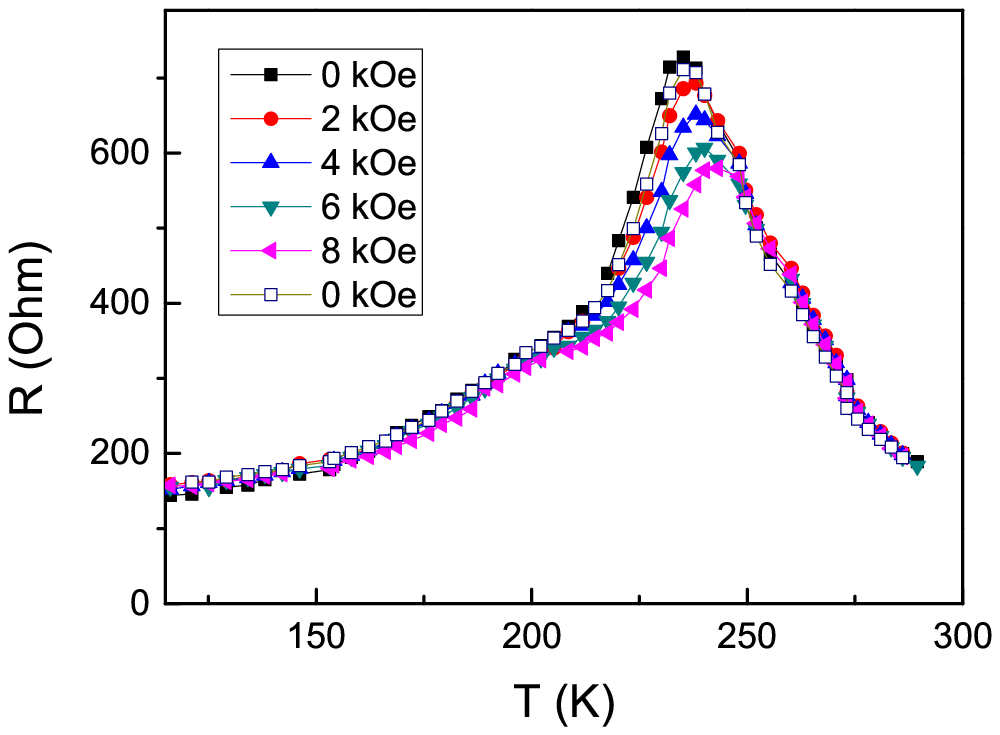}
\caption{Temperature dependence of the resistance  of the heterostructure in different magnetic fields applied perpendicular to the interface (and parallel to the $c$-axis of the ferromagnetic state of LaMnO$_3$) after many cycles of the external magnetic field.}
\end{figure}
As it is seen from this figure the field dependence of R(T) curves still
exists but it is reversible. Initial values of resistance restores
after switching off the external magnetic field. The origin of this
magnetoresistance differs considerably from the observed above. It
is reversible and can be attributed to the magnetoelastic effect.

 The final step of our study is the application of the external magnetic field
 perpendicular to the c-axis. The results of this study are presented in Fig.~5.
\begin{figure}[h]
\includegraphics[width=0.9\columnwidth,angle=0]{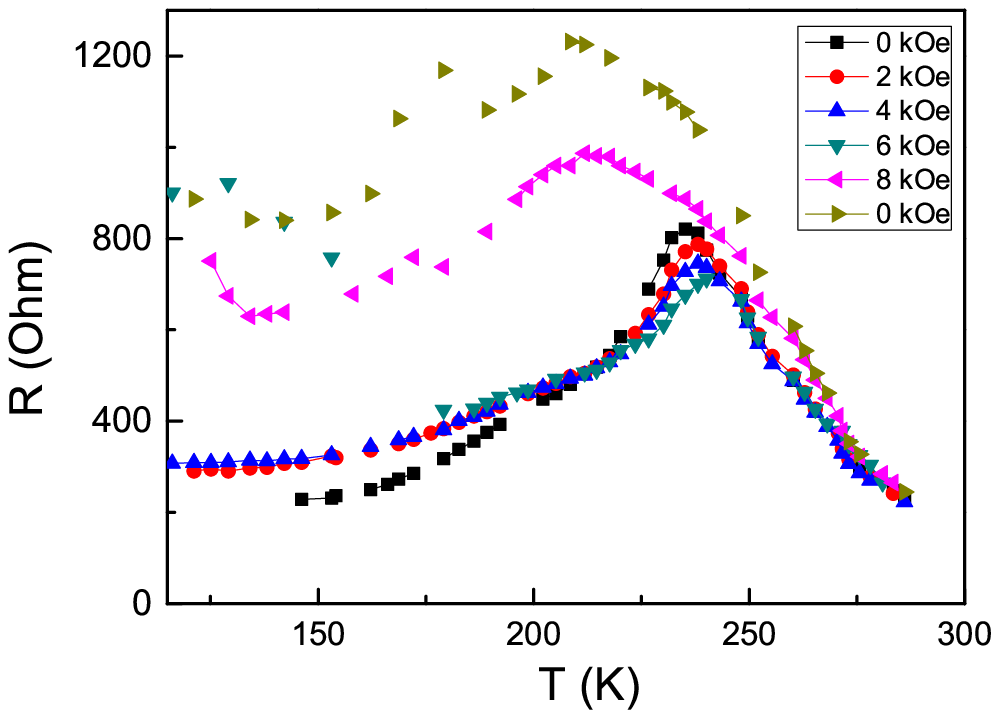}
\caption{Temperature dependence of the resistance  of the heterostructure in different magnetic fields applied parallel to the interface (and perpendicular to the $c$-axis of the ferromagnetic state of LaMnO$_3$).}
\end{figure}
As it is seen from this figure, the temperature dependence of the
resistance doesn't depend on the external magnetic field up to 6
kOe. Further increase of the field leads to the distortion of the
$R(T)$ dependence. Low temperature wing of the $R(T)$ curve start to
rise.

The character of the temperature dependence of the resistance
reveals the appearance of metallic contribution $R\sim T$ to the
total resistance of the Ba$_{0.8}$Sr$_{0.2}$TiO$_3$/LaMnO$_3$ heterostructure. The magnitude of this
contribution is comparable with the total
resistance of LaMnO$_3$. Bearing in mind \cite{Pavlov} that the
density of the 2DEG is extremely large up to (1.7 -
3.0){$\times 10^{14}$ cm$^{-2}$, one can expect that
the interface area of the antiferromagnetic LaMnO$_3$ is strongly
modified. Such observation is the most important result of the
present paper.

As we know from our previous results \cite{Pavlov}, a high-mobility 2DEG arises within few
atomic layers of the Ba$_{0.8}$Sr$_{0.2}$TiO$_3$/LaMnO$_3$ interface. The most common mechanism for 2DEG is the polar discontinuity model~\cite{Biscaras}, which was also discussed for the case of ferroelectric/dielectric interface~\cite{Pavlov,c22,c24,c25,c26}. The polar discontinuity at the interface leads to the divergence of the
electrostatic potential.  In order to minimize the total energy, it
is necessary to shield the electric field arising from this. As a
consequence, both the lattice system and the energy spectrum of the
current carriers are restructured~\cite{c22}, and the increase of
the current carriers density occurs in a narrow interface area. This
occurs in a self-consistent manner, so that rearranging the energy
spectrum of the carriers and increasing their concentration in the
interface region leads to the formation of a narrow metal region
near the interface. If we assume that the spontaneous polarization of the Ba$_{0.8}$Sr$_{0.2}$TiO$_3$ ferroelectric is $P_S \approx$ 30 $\mu$K/cm$^2$, it gives the surface charge carrier density $n_S$=1,875\ 10$^{14}$ cm$^{-2}$.
This concentration allows to estimate the mobility of carriers in this highly condacting layer on the interface. Assuming that the thickness of the 2DEG is $\Delta_S$ we may estimate the 3D resistivity of the layer $\rho=R_S\Delta_S y$, where $y=L/W$ ($L$ is the length of the section across which the resistance is measured, and $W$ is the transverse width of the resistance measurements, $y$= 0,5-1 in our case), and and $R_S$ is resistance of heterostructure at 100 K. Therefore, assuming that 3D charge carrier concentration in the interface layer is $n_S/\Delta_S$ we obtain the estimate for the mobility $\mu=y/R_Sen_s$ ($e$ is elementary charge). Note that this estimate does not depend on the conducting layer thickness $\Delta_S$. Our estimates give $\mu\approx$ 10 $cm^2/V s$ for poorly conducting metallic state and $\mu\approx $ 100 $cm^2/V s$ for highly conductive metallic state after magnetic field application.
The conducting layer thickness may be estimated assuming that 3D conductivity in this layer is equal to the conductivity in La$_{x}$Sr$_{x}$MnO$_3$ ($x$ = 0.17-0.30). This yields $\Delta_S=\rho_v y/ R_S$, where $\rho_v$ is the electrical resistivity of manganite in the metallic state ($\rho_v = $6 10$^{-5}$- 2 10$^{-4}$ Ohm$\cdot$cm for La$_{x}$Sr$_{x}$MnO$_3$ with $x$ = 0.17-0.30 at T=100 K). As a results, we obtain $\Delta_S \approx $ 1,5-10 nm.

De Gennes \cite{Gennes} has analyzed
the effect of the appearance of the so-called Zener carriers
\cite{Zener} in an antiferromagnet, which leads to a formation of a
lower energy ferromagnetic configuration. In particular, he
has considered the situation of the
antiferromagnetic LaMnO$_3$ manganite. This "double exchange" is
completely different from the usual indirect RKKY exchange. The
main property of the double exchange coupling is that it
destroys the antiferromegnetic spin order.

Our results are related to the fundamental physical transformation
of the magnetic state under the influence of the high density
current carriers at the interface. We suppose that before the
application of the external magnetic field, the ferromagnetic
regions with different orientations of magnetic moments are formed. Applying the
magnetic field perpendicular to the interface (and
parallel to the easy {\it c} axis of LaMnO$_3$) leads to decrease of
the resistance and shifting of the maximum of $R(T)$ to the
high-temperature side. After several
%
%
cycles of similar manipulations by an external field the
temperature behavior of the resistance is stabilized (Fig. 4) with
the maximum at $T$=240 K and is not changed. Only a small
magnetoresistive effect is observed (Fig. 3).
The application of a magnetic field perpendicular to the {\it c}
axis of LaMnO$_3$ manganite ({\it c} axis is the easy magnetization
axis) leads to the degradation of the highly conductive state. — The
resistance in the low-temperature region increases strongly and does
not show metallic behavior (see Fig. 5).

 This behavior has natural explanation.
 We suppose that the observed variation of the electrical
resistance is cased by the formation of the ferromagnetic
order. \cite{Dagotto,Mukhin,Mukhin2,Mamin07,
Mamin07J}. At the first stage of the modification of the
ferromagnetic state the ferromagnetic regions with different orientations of magnetic moments are formed. This leads to a relatively small resistance. Under the
application of the magnetic field the magnetic structure starts to
reconstruct and the larger {in size ferromagnetic regions with uniform
magnetization are formed. As a result, the number of the scattering
centers at the domain boundaries decreases, and the electrical
resistance decreases. When a whole interface region forms
homogeneously magnetized domain the "final" temperature dependence
of the resistance with the maximum at $T$=240 K is observed.
When the magnetic field is applied perpendicular to the {\it c}-axis
of the LaMnO$_3$ single crystal, strong magnetostriction effects are
expected in the ferromagnetic phase of the system. The magnetic
field tends to orient the magnetization in the interface region
along the field. We believe that strong stresses and distortions
occur in the interface area. These very strong distortions in this
very thin layer, may partially mechanically destroy the the
interface area. This leads to  substantial and irreversible
increases of the resistance of the interface. As a result, all or
part of the high-conductive region is lost, and the resistance
strongly increases at low temperatures.

In our investigation, we have found the creation of a magnetic state
in LaMnO$_3$ due to the proximity effect with a ferroelectric. The
ferroelectric state may be controlled by an electric field,
 therefore there is a possibility for turning ferromagnetic state off and
on by the external electric field, which changes the direction of
polarization in the ferroelectric film.
This allows us to control the interface magnetic state by applying
an electric field, as it was done in the case of interface
superconductivity tunable by electric field applied directly to
the interface~\cite{Th,Tri} or by electric field applied through an
ionic
 liquid~\cite{Gold}. Therefore, it opens the possibility to use
these phenomena in the design of novel electronic devices.

We have studied the electrical properties of the heterostructure 
of the ferroelectric Ba$_{0.8}$Sr$_{0.2}$TiO$_3$ film
deposited on the single crystalline manganite LaMnO$_3$ under the
influence of the external magnetic field. We obtained the indirect
evidence for the formation of small ferromagnetic regions near the
interface.


We gratefully acknowledge I.~A.~Garifullin for fruitful discussion.
This work was supported by Russian Foundation for Basic Research
through project
 No.18-12-00260. V.V.K. acknowledges financial support from Slovenian Research Agency Program P1-0040.


\begin{thebibliography}{99}

\bibitem{Bozovic} Ivan Bozovic, IEEE Trans. Appl. Superc. {\bf 11}, NO. I , (2001).


\bibitem{Ohtomo} A. Ohtomo and H. Y. Hwang, Nature 427, {\bf 423}, (2004).

\bibitem{Th}
S.~Thiel, G.~Hammerl, A.~Schmehl, C.~W.~Schneider, and J.~Mannhart,
Science {\bf 313}, 1942 (2006).

\bibitem{Rey}
N.~Reyren, S.~Thiel, A.~D.~Caviglia, L.~Fitting~Kourkoutis,
G.~Hammerl, C.~Richter, C.~W.~Schneider, T.~Kopp,
A.-S.~R{\"u}etschi, D.~Jaccard,M.~Gabay, D.~A.~Muller,
J.-M.~Triscone, and J.~Mannhart, Science {\bf 317}, 1196 (2007).

\bibitem{Bri}
A.~Brinkman, M.~Huijben, M.~Van Zalk, J.~Huijben, U.~Zeitler,
J.~C.~Maan, W.~G.~van der Wiel, G.~Rijnders, D.~H.~A.~Blank, and
H.~Hilgenkamp, Nature Materials~{\bf 6}, 493 (2007).

\bibitem{Moetakef} P. Moetakef, T. A. Cain, D. G. Ouellette, J. Y. Zhang, D. O.
Klenov, A. Janotti, Ch. G. van de Walle, S. Rajan, S. J. Allen, and
S. Stemmer, Appl. Phys. Lett. {\bf 99}, 232116 (2011).

\bibitem{Jackson} C. A. Jackson and S. Stemmer, Phys. Rev. B 88, 180403(R) (2013).

\bibitem{Biscaras} J. Biscaras, N. Bergeal, A. Kushwaha, T. Wolf, A. Rastogi, R. C.
Budhani, and J. Lesueur, Nat. Commun. {\bf 1}, 89 (2010).

\bibitem{Annadi} A. Annadi, A. Putra, Z. Q. Liu, X. Wang, K. Gopinadhan, Z.
Huang, S. Dhar, T. Venkatesan, and Ariando, Phys. Rev. B {\bf 86},
085450 (2012).

\bibitem{Perna} P. Perna, D. Maccariello, M. Radovic, U. Scotti di Uccio, I.
Pallecchi, M. Codda, D. Marré, C. Cantoni, M. Varela, S. J.
Pennycook, and F. M. Granozio, Appl. Phys. Lett. {\bf 97}, 152111
(2010).

\bibitem{Kal}
A.~Kalabukhov, R.~Gunnarsson, J.~B{\"o}rjesson, E.~Olsson,
T.~Claeson, and D.~Winkler, Phys.~Rev.~B {\bf 75}, 121404(R) (2007).

\bibitem{Chen}
Y.~Chen, F. Trier, T. Kasama, D. V. Christensen, N. Bovet, Z. I.
Balogh, H. Li, K. T. S. Thyden, W. Zhang, S. Yazdi, P. Norby, N.
Pryds, S.~Linderoth, Nano Lett. {\bf 15}, 1849 (2015).

\bibitem{Pavlov} D.P. Pavlov, I. I. Piyanzina, V. I. Muhortov,
A.I. Balbashov, D. A. Tauyrskii, I. A. Garifullin, R. F. Mamin, JETP Lett.{\bf 106}, 460 (2017).

\bibitem{Pavlov2} D. P. Pavlov, R. R. Zagidullin, V. M. Mukhortov, V. V. Kabanov,
T. Adachi, T. Kawamata, Y. Koike, and R. F. Mamin, Phys. Rev. Lett. {\bf 122}, 237001 (2019).

\bibitem{Dagotto} E. Dagotto, T. Hotta, and A. Moreo, Phys. Rep. {\bf 344}, 1 (2001).

\bibitem{Gennes} P.-G. de Gennes, Phys. Rev. {\bf 118}, 141 (1960).

\bibitem{Zener} C. Zener, Phys.~Rev. {\bf 82}, 403 (1951).

\bibitem{Mukhortov1} V.M. Mukhortov, Y.I. Golovko, G.N. Tolmachev, A.N. Klevtzov, Ferroelectrics {\bf 247}, 75 (2000).

\bibitem{Mukhortov2} V. M. Mukhortov, G. N. Tolmachev, Yu. I. Golovko, and A.I. Mashchenko, Tech. Phys. {\bf 43}, 1097 (1998). 

\bibitem{Mukhortov3}  V. M. Mukhortov, Yu. I. Golovko, G. N. Tolmachev, and A.I. Mashchenko, Tech. Phys. {\bf 44}, 1477 (1999).

\bibitem{Mukhin}A. A. Mukhin, V. Yu. Ivanov, V. D. Travkin, S. P. Lebedev,
A. Pimenov, A. Loidl, and A. M. Balbashov, JETP Lett. {\bf 68}, 356 (1998).

\bibitem{Mukhin2}
V. Yu. Ivanov, V. D. Travkin, A. A. Mukhin, S. P. Lebedev, A. A.
Volkov, A. Pimenov, A. Loidl, A. M. Balbashov and A.V. Mozhaev, J. Appl. Phys.
\textbf{83}, 7180 (1998).

\bibitem{c22} V. V. Kabanov, I. I. Piyanzina, D. A. Tayurskii, and R. F. Mamin, Phys. Rev. B \textbf{98}, 094522 (2018).
\bibitem{c24} M.K. Niranjan, Y. Wang, S.S. Jaswal, and E.Y. Tsymbal, Phys. Rev. Lett. \textbf{103}, 016804 (2009).

\bibitem{c25} Y. Wang, M.K. Niranjan, J.D. Burton, J.M. An, K.D. Belashchenko, E.Y. Tsymbal, Phys. Rev. B \textbf{79}, 212408 (2009).

\bibitem{c26} X. Liu, E. Y. Tsymbal, K. M. Rabe, Phys. Rev. B \textbf{97}, 094107 (2018).


\bibitem{Mamin07} R.~F.~Mamin, T.~Egami, Z.~Marton, and S.~A.~Migachev, Phys. Rev. B \textbf{75}, 115129 (2007).


\bibitem{Mamin07J} R.~F.~Mamin, T.~Egami, Z.~Marton, S.~A.~Migachev, and M.~F.~Sadyikov, JETP Lett. \textbf{86}, 643 (2007).


\bibitem{Tri}
A. D. Caviglia, S. Gariglio, N. Reyren, D. Jaccard, T.~Schneider, M.
Gabay, S. Thiel, G. Hammerl , J. Mannhart and J.-M. Triscone Nature
{\bf 456}, 624 (2008).

\bibitem{Gold}
X.~Leng, J.~Garcia-Barriocanal, S.~Bose, Y.~Lee, and A.~M.~Goldman,
Phys.~Rev.~Lett. \textbf{107}, 027001 (2011).


\end{thebibliography}
\end{document}